\documentclass[aps,prl,twocolumn,groupedaddress,nofootinbib,showpacs]{revtex4}
\usepackage[dvips]{graphicx}
\usepackage{amsmath,amssymb,slashed}
\newif\ifdraft
\drafttrue
\newif\ifpreprint
\preprinttrue

\def\spa#1.#2{\left\langle#1\,#2\right\rangle}
\def\spb#1.#2{\left[#1\,#2\right]}

\newcommand{\eq}{\begin{equation}}
\newcommand{\eqe}{\end{equation}}
\newcommand{\eqa}{\begin{eqnarray}}
\newcommand{\eqae}{\end{eqnarray}}

\newbox\charbox
\newbox\slabox
\def\s#1{{      
        \setbox\charbox=\hbox{$#1$}
        \setbox\slabox=\hbox{$/$}
        \dimen\charbox=\ht\slabox
        \advance\dimen\charbox by -\dp\slabox
        \advance\dimen\charbox by -\ht\charbox
        \advance\dimen\charbox by \dp\charbox
        \divide\dimen\charbox by 2
        \raise-\dimen\charbox\hbox to \wd\charbox{\hss/\hss}
        \llap{$#1$}
}}

\begin{document}

\title{
\ifpreprint
\hbox{\rm \small 
$\null$} 
\fi
New fermionic soft theorems for supergravity amplitudes}
 
\author{Wei-Ming Chen$^{a}$, Yu-tin Huang$^{a,b}$, Congkao Wen$^c$
}

\affiliation{
${}^a$Department of Physics and Astronomy, National Taiwan University, Taipei 10617, Taiwan, ROC\\
${}^b$ School of Natural Sciences, Institute for Advanced
Study, Princeton, NJ 08540, USA\\
${}^c$Dipartimento di Fisica, Universit\`a di Roma ``Tor Vergata" \& I.N.F.N. Sezione di Roma ``Tor Vergata", Via della Ricerca Scientifica, 00133 Roma, Italy}

\begin{abstract} 
Soft limits of massless S-matrix are known to reflect symmetries of the theory. In particular for theories with Goldstone bosons, the double-soft limit of scalars reveals the coset structure of the vacuum manifold. In this letter, we propose that such universal double-soft behavior is not only true for scalars, but also for spin-1/2 particles in four dimensions and fermions in three dimensions. We first consider Akulov-Volkov theory, and demonstrate the double-soft limit of Goldstinos yields the supersymmetry algebra. More surprisingly we also find amplitudes in $4 \leq \mathcal{N} \leq 8$ supergravity theories in four dimensions as well as $\mathcal{N}=16$ supergravity in three dimensions behave universally in the double-soft-fermion limit, analogous to the scalar ones. The validity of the new soft theorems at loop level is also studied. The results for supergravity are beyond what is implied by SUSY Ward identities, and may impose non-trivial constraints on the possible counter terms for supergravity theories.
\end{abstract}

\pacs{04.65.+e, 11.15.Bt, 11.30.Pb, 11.55.Bq \hspace{1cm}}

\maketitle

The connection between symmetries of a theory and the soft behavior of its S-matrix has been previously explored in various examples. The most famous case is Weinberg's soft graviton theorem~\cite{Weinberg}, which states that the leading soft divergence of a gravitational S-matrix is constrained by Ward identities and is  universal. Similar results, based on other symmetries, were shown to imply universality for the subleading divergences in gauge and gravity theories as well~\cite{Low,SoftGravy1, CS, BernGauge, SymArg}. Another famous example, which is more relevant to this letter, is in the context of the Goldstone bosons of a spontaneous broken symmetry. In particular, taking the momentum of the Goldstone boson to near zero, which corresponds to a constant scalar field, the S-matrix should vanish due to the scalars being derivatively coupled. This is the well-known Adler's zero~\cite{AdlerZero, Weinberg:1966kf}.

As discussed in~\cite{Simplest}, Adler's zero can also be understood from the structure of the vacuum. Perturbative scattering amplitudes should be identical when computed at any point in the vacuum moduli. On the other hand, one can in principle use the operator $e^{i\theta \cdot T}$ to relate one vacuum to another via $|\theta\rangle=e^{i\theta \cdot T}|0\rangle$, where $T^a$ represents the broken generators, and $\theta_a$ is a constant that is the vacuum expectation value (vev) of the soft scalar. The amplitude evaluated in the $|\theta\rangle$ vacuum can be written as a perturbative series by expanding out the exponent, 
\eq
|\theta\rangle=e^{i\theta T}|0\rangle=|0\rangle+\theta_i|\pi^i\rangle+\frac{1}{2}\theta_i\theta_j|\pi^i\pi^j\rangle+\cdots\,,
\eqe 
where $|\pi^i\rangle, |\pi^i\pi^j\rangle$ e.t.c represent vacuums with one and two (or more) additional soft scalars, respectively. Since the amplitude is the same in either $|\theta\rangle$ or $|0\rangle$ vacuum, this implies that amplitudes with one or more soft scalars must vanish. The vanishing of the single soft scalar is precisely Adler's zero. For two soft scalars, it turns out that the amplitude is non-zero due to the non-commutativity of the broken generators. It was shown in~\cite{Simplest} that taking two Goldstone bosons to have soft momenta, the amplitude behaves as:
\eqa\label{GenSoft}
\hspace{-0.3mm} \notag M_{n}\hspace{-1.1mm}\left[\phi^i (t^2 p_1),\phi^j (t^2 p_2)\cdots \right]{\hspace{-0.1cm}\bigg|_{t\rightarrow 0}}\hspace*{-0.3cm}=\sum_{a=3}^n \mathcal B_a f^{ijK}{H_a}_K M_{n{-}2}\,,\\[-0.85cm]~\\[-0.3cm] \notag 
\eqae
where $\mathcal B_a\equiv \frac{p_a\cdot (p_1-p_2)}{2p_a\cdot (p_1+p_2)}$ and ${H_a}_K$ is the generator of the invariant subgroup in the coset $G/H$, while $f^{ijK}$ is the structure constant in $[T^i, T^j] =f^{ijK}H_K$.

In extended supergravity theories the scalars also parameterize a coset space, and the double-soft-scalar limit of the S-matrix is given by~eq.(\ref{GenSoft}) as well. The fact that the non-linearly realized symmetries have non-trivial imprint on the S-matrix is extremely useful in the discussion of ultra-violet behavior of supergravity theories. In particular, modulo quantum anomalies, any possible counter terms for the theory must respect this symmetry, which can be verified by checking whether or not the S-matrix elements generated by such counter terms agree with eq.(\ref{GenSoft})~\cite{ElvangR4, ElvangFull}.    

In this letter, we demonstrate that remarkably, the same single and double-soft behavior also applies to fermions in Akulov-Volkov (A-V) theory~\cite{VAModel} as well as supergravity theories both in three and four dimensions. For A-V theory, which is the effective action for the Goldstinos of spontaneously broken supersymmetry, we show that the amplitudes vanish in the single-soft-fermion limit. In the double-soft limit they exhibit a similar form as that of the scalars in~eq.(\ref{GenSoft}) with $H_{aK}$ replaced by $[1|p_a|2\rangle$, where legs 1 and 2 are $+1/2$ and $-1/2$ helicity fermions respectively and with all momenta outgoing.\footnote{We use standard spinor-helicity formalism in four and three dimensions. For 4D: $p^{\alpha \dot{\alpha}}_i = \lambda^{\alpha}_i \tilde{\lambda}^{\dot{\alpha}}_i$, with scalar products as $\lambda_i^\alpha\lambda_j^\beta\epsilon_{\alpha\beta}=\langle ij\rangle$, $\tilde\lambda_{i\dot\alpha}\tilde\lambda_{j\dot\beta}\epsilon^{\dot\alpha\dot\beta}=[ij]$, $s_{ij}=\langle ij\rangle[ji]$, and $[i|p_a|j\rangle = [i a] \langle a j \rangle$. In Minkowski signature, $\tilde{\lambda}=\lambda^*$. Reduction to three-dimensions simply corresponds to $\lambda$ being real.} The latter precisely reflects the supersymmetry algebra, $\{Q,\bar{Q}\}=P$, with additional factors proportional to the external-line factors of the soft fermions. 

For the supergravity theories, we show that the amplitudes vanish in the single-soft limit of spin-$1/2$ particles in four dimensions as well as all fermions in three dimensions. In the double-soft-fermion limit, amplitudes in four-dimensional supergravity theories again behave in an analogous way as that of the scalars in~eq.(\ref{GenSoft}), now with: $p_a{\cdot} (p_1{-}p_2){\rightarrow} [1|p_a|2\rangle$. Notice the recurrence of the factor $[1|p_a|2\rangle$, as with the A-V theory. In three dimensions, we show that for $\mathcal{N}=16$ supergravity~\cite{E8}, where the 128 scalars parametrize the E$_{8(8)}$/SO(16) coset space, the double-soft limit of any pair of the $128$ fermions exhibits the same behavior as the spin-$1/2$ fermions in four dimensions. 
\section{Double-soft limit and spontaneous (super)symmetry breaking} 
Scattering amplitudes involving Goldstone bosons have interesting soft behavior that reveals the details of the vacuum. In~\cite{Simplest} it was shown that for $\mathcal{N}=8$ supergravity, where the 70 scalars parameterize E$_{7(7)}$/SU(8), the amplitude with two soft scalars behaves as: 
\eqa\label{GenSoft1}
 &&\hspace{-0.5cm}M_{n}\left(\phi^{I_1I_2I_3I_4}(t^2 p_1),\,\phi^{I_5I_6I_7I_8} (t^2 p_2)\cdots \right)\bigg|_{t\rightarrow 0}\\[-1.2mm]
\notag &&= 4\sum_{a=3}^n\mathcal B_a\epsilon^{I_1I_2I_3I_4[I_5I_6I_7| J}(R_a)^{I_8]}\,_J M_{n{-}2}{+}\mathcal{O}(t^2)\,,
\eqae
where $I_i, J_i=1,\cdots 8$ are the SU(8) R-symmetry indices, and the square bracket $[\,\,\,]$ indicates anti-symmetrization. The generators $(R_a)^{I}\,_J$ in eq.(\ref{GenSoft1}) are the single site SU(8) generators, where $a$ labels the external leg. The soft limit $p_{1,2} \rightarrow t^2 p_{1,2}$, is realized on the spinors as 
\eqa\label{Softlimit}
\lambda_{1,2} \rightarrow t \lambda_{1,2}\,, \quad \tilde{\lambda}_{1,2} \rightarrow t \tilde{\lambda}_{1,2}\, .
\eqae
This analysis was later extended to $D=4$, $4\leq\mathcal{N}<8$, and $D=3$, $\mathcal{N}=16$ supergravity in~\cite{US}. One new subtlety is the presence of U(1) factors in the isotropy group $H$, which produces soft-graviton singularities in the double-soft limit. Such singularities are absent in the $\mathcal{N}=8$ theory. To extract the finite piece one instead considers the anti-symmetrized double-soft limit $M^{[i,j]}_{n}$, defined as:
\eq
\label{anti}
M^{[i,j]}_{n}\equiv M_n\left( \phi^i( t^2 p_1),\; \phi^j( t^2 p_2),\cdots,n \right)-(1\leftrightarrow 2)\bigg|_{ t {\rightarrow}0}\,.
\eqe
Indeed for all $4\leq\mathcal{N}<8$ supergravity in four dimensions and $\mathcal{N}=16$ supergravity in three dimensions, the anti-symmetrized double-soft limit yields:
\eq
M^{[i,j]}_{n}=\sum_{a=3}^n\frac{p_a\cdot (p_1-p_2)}{p_a\cdot (p_1+p_2)}f^{ijK}(H_a)_K M_{n{-}2}{+}\mathcal{O}(t^2)\,,
\eqe
where the $(H_a)_K$s are single site U($\mathcal{N}$) generators for $D{=}4$, $4<\mathcal{N}<8$, U(1) for $D{=}4$ $\mathcal{N}=4$ and SO($\mathcal{N}$) for $D{=}3$.

Another interesting example we would like to consider is the A-V theory~\cite{VAModel}, which is the low energy effective action of fermions (Goldstinos) associated to spontaneous breaking of supersymmetry. The action is given by
\eqa
S_{\rm AV}=-\frac{1}{2 g^2}\int d^4x \det (1+ig^2\psi \sigma^\mu\overset{\leftrightarrow}{\partial}_\mu \bar{\psi} )
\, ,
\eqae
where the Weyl fermion $\psi$ is the Goldstino and $\sigma^\mu{=}(1,\overset{\rightarrow}{\sigma})$, with $\overset{\rightarrow}{\sigma}$ as the Pauli matrices.
One can expand the determinant which generate an infinite series of local operators. Here we will use the six-point amplitude as an example. The relevant vertices are the four-point vertex:
{\hspace*{-0.47cm}\begin{minipage}{1\textwidth}
\begin{minipage}[c]{0.01\textwidth}
\hspace{-1.8cm}\includegraphics[scale=0.6]{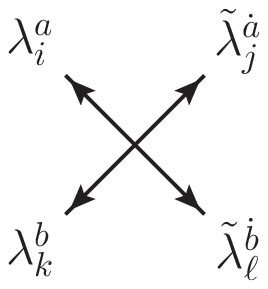}~~~
\end{minipage}~~~~~~~~~~~\hspace{-0.07cm}\begin{minipage}[l]{0.43\textwidth}
\vspace*{-0.3cm}\eqa
\notag&&g^2[V_{jk}
-V_{kj}]_{ab\dot a\dot b}-(i,_a\leftrightarrow k,_b)-(j,_{\dot a}\leftrightarrow \ell,_{\dot b})\\
&&\notag~~~~~~~~~~~~~~~~~~~~+[(i,_a,j,_{\dot a})\leftrightarrow (k,_{b},\ell,_{\dot b})]\,,\\
\eqae
\end{minipage}
\end{minipage}},\\
where $V_{ij,ab\dot a\dot b}\equiv (p_i)_{a\dot a}(p_j)_{b\dot b}$, and the six-point vertex:\\
{\hspace*{-0.47cm}\begin{minipage}{1\textwidth}
\begin{minipage}[c]{0.01\textwidth}
\hspace{-1.8cm}\includegraphics[scale=0.6]{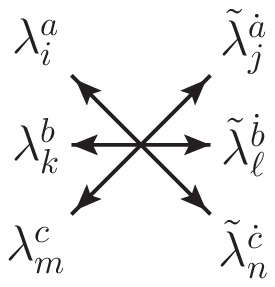}~~~
\end{minipage}~~~~~~~~~~~\hspace{-0.1cm}\begin{minipage}[l]{0.43\textwidth}
\eqa
\notag&&\sum_{\sigma\in\mathrm{perm.}}(-)^\sigma V_6(\sigma(i,_a,j,_{\dot a},k,_b,\ell,_{\dot b},m,_c,n,_{\dot c}))\,,\\ 
\eqae
\end{minipage}
\end{minipage}}\\
where \\[1.5mm]
$\begin{array}{cl}
& V_6(i,_a,j,_{\dot a},k,_b,\ell,_{\dot b},m,_c,n,_{\dot c}) \equiv i\frac{g^4}{4}[2V_{\ell m j}-2V_{jm\ell}\cr
 &-V_{\ell jm}+V_{j\ell m}+2V_{kni}+2V_{ink}+V_{kin}-V_{ikn}]_{abc\dot a\dot b\dot c}\,,
\end{array}$ \\[1.5mm]
with $V_{ijk,abc\dot a\dot b\dot c}\equiv (p_i)_{a\dot a}(p_j)_{b\dot b}(p_j)_{c\dot c}$. A straightforward computation shows that the relative coefficients between the quartic and sextic interactions are precisely needed for the leading terms in the single-soft limit to cancel, such that the six-point amplitude is of order $t^2$ in the limit.  In the double-soft limit one finds:\\[-3mm]
\eqa
&&\hspace*{-3mm}M_6(\psi_1,\bar{\psi}_2,\psi_3,\bar{\psi}_4,\psi_5,\bar{\psi}_6)|_{\lambda_{1,2}\rightarrow t\lambda_{1,2}\atop{\tilde{\lambda}_{1,2}\rightarrow t\tilde{\lambda}_{1,2}}}\\[-2mm]
\notag&&~~~~~~=t^2g^2\sum_{a=3}^6 \mathcal B_a [1| p_a |2 \rangle M_4(\psi_3,\bar{\psi}_4,\psi_5,\bar{\psi}_6)+\mathcal{O}(t^4)\,,
\eqae
where
 $M_4(\psi_3,\bar{\psi}_4,\psi_5,\bar{\psi}_6)=2g^2 s_{46} [35]\langle 46\rangle$. The above results are consistent with the interpretation that the fermions are Goldstinos. From the single- and double-soft behaviors of Goldstone bosons, one would expect that the single-soft limit of a Goldstino should vanish as $\mathcal{O}(t)$, while the double-soft limit should be finite and proportional to $\mathcal{B}_ap_aM_{n{-}2}$, due to the fact that the broken generators are associated with $Q$, and satisfy the algebra $\{Q,\bar{Q}\}=P$. The extra factor of $t$ for the single-soft limit, and $t^2\lambda_1\tilde{\lambda}_2$ for the double-soft case, are simply due to the presence of additional soft external-line factors for fermions.
\section{New double-soft theorems in four dimensions }  
We now consider soft fermions in four-dimensional supergravity theories. As discussed in ref.~\cite{US}, due to the fact that amplitudes with a soft scalar vanish as $\mathcal{O}(t^2)$, SUSY Ward identities~\cite{SUSYWard} require that the amplitudes with a soft fermion vanish as $\mathcal{O}(t)$. The same result can alternatively be deduced from BCFW recursion~\cite{BCFW}. However, for the double-soft limit, Ward identities are no longer sufficient since it implies:
\eq
\begin{array}{rcl}
0 &=& \langle 0| [Q, \phi_1 \psi_2, \ldots ]  |0 \rangle
\cr
&=& [1 q]M(\psi_1 \psi_2, \ldots) + [2 q] M(\phi_1 \phi_2, \ldots) \cr
&&+ 
\sum_{i>2} [iq] M(\phi_1 \psi_2, \ldots)  \,,
\end{array}
\eqe~\\[-2mm]
where $q$ is a reference momentum. As one can see the identity yields a linear relation between double-soft fermions, scalars and soft-scalar-fermion limits. Thus the double-scalar limits by itself is insufficient to determine the double-fermion limits. Instead we will proceed using the recursion relations.

To treat all the hard particles democratically, we add an auxiliary negative-helicity graviton~\cite{Simplest}, which at the end is taken away by sending its momentum to be soft\footnote{This trick is of course not necessary, a different derivation without the auxiliary graviton is presented in~\cite{US}}. Now, we choose the shifted legs in the recursive formula as one of the soft legs and the added graviton. The remaining soft leg will be in one of the factorized amplitudes which generally vanishes due to the previous analysis. Thus all diagrams vanish except for the following two special cases: ~\\[-0.6cm]
\eqa
\nonumber\vcenter{\hbox{\includegraphics[scale=0.45]{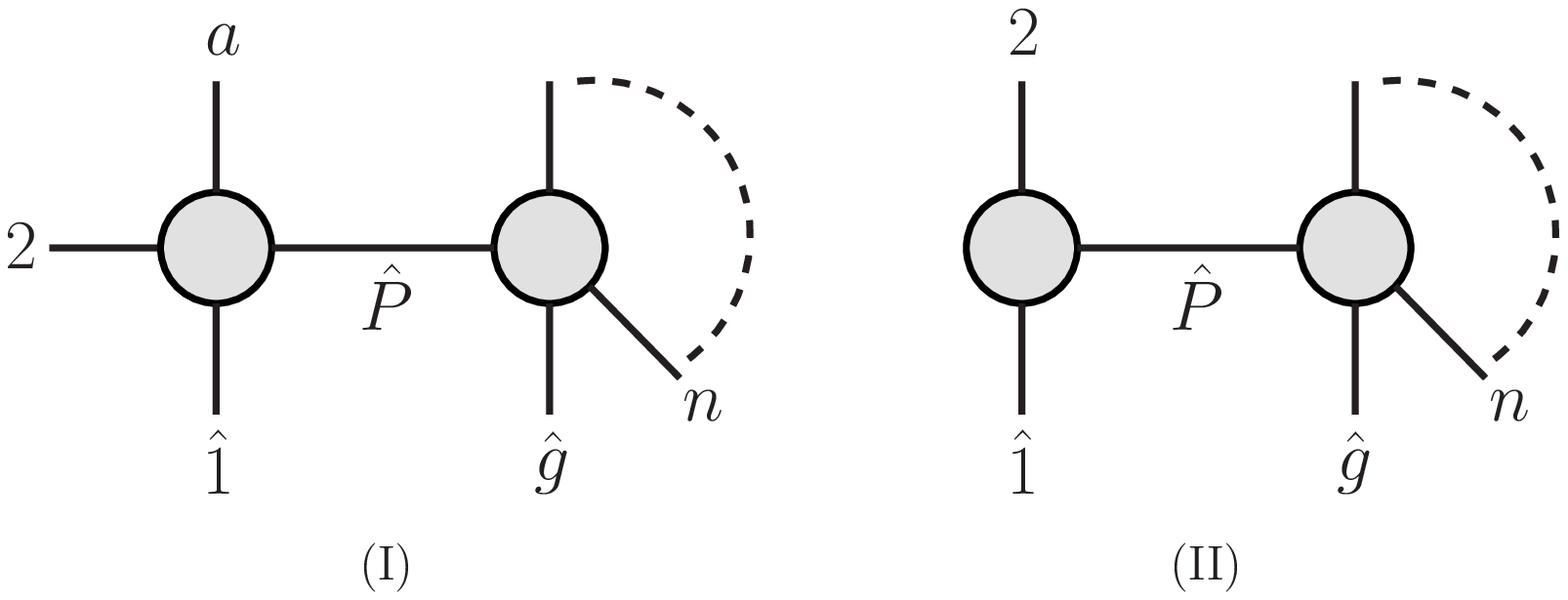}}}\,.
\eqae~\\[-0.5cm]
where the hatted labels indicate they are the deformed legs, and must be at opposite side of the propagator for any diagram in the recursion. Unlike the scalars for which one can use the anti-symmetric extraction scheme in eq.(\ref{anti}) to obtain the U(1) part of the symmetry group~\cite{US}, here U(1) is inaccessible due to the soft-fermions having opposite helicities and thus cannot be ``symmetrized". Thus we will consider the case where two fermions do not form a singlet, for which only the diagram (I) contributes, and to the leading order it is given as~\cite{US} (We have presented the $\mathcal{N}=8$ result, from which lower superymmetric theories can be obtained via SUSY truncation~\cite{LessSUSY} ): 
\eqa\label{Initial}
 &&\hspace{-1cm} \mathcal{M}_4(\hat{1}, 2, a, \hat{P})\langle \hat{1}P\rangle^{8}\frac{ \delta^8\left(\eta_1+\frac{\langle {P} 2\rangle}{\langle {P} \hat{1}\rangle}\eta_2+ \frac{\langle {P} a\rangle}{\langle {P} \hat{1}\rangle}\eta_a \right) }{2 p_a\cdot (p_1+p_2) }\\
\notag 
&&~~~~~~~~~~~~~~~\times \exp{\left(-\frac{\langle \hat{1}2\rangle}{\langle \hat{1}P\rangle}\eta^I_2\frac{\partial}{\partial\eta^I_a}\right)}  \mathcal{M}_{n{-}1}(g)\,.
\eqae
Here the $\mathcal{M}_i$'s are superamplitudes, and $\mathcal{M}_{n{-}1}(g)$ indicates one of the legs being a soft graviton.  The superamplitudes are homogenous polynomial function of the $\eta^I$s whose coefficients are the component amplitudes. The $\eta^I$s are Grassmann odd variables carrying fundamental SU(8) indices. They are used to package all on-shell degrees of freedom into a superfield, which for $\mathcal{N}=8$ is given by:
\eqa
\Phi(\eta) = G^{{++}} {+} \eta^A \Lambda_A {+} { \eta^A \eta^B \over 2!}  A^+_{AB}{+} \ldots {+} (\eta)^8 G^{{--}} \,,
\eqae
where $G^{++}$ is the $+2$ graviton, $\Lambda_A$ is the $+{3 \over 2}$ gravitino, and $A^+_{AB}$ the $+1$ graviphoton and so on. The hatted labels in eq.(\ref{Initial}) indicate the corresponding legs are deformed due to the BCFW shift, and their explicit forms can be found in~\cite{US}. 
For the fermion pair $(\psi_1^{I_1I_2I_3},\bar{\psi}_2^{I_4I_5I_6I_7I_8})$, diagram (I) yields: 
\eqa \label{diagramI}
\nonumber&&\frac{5[1|p_a |2\rangle}{ 2p_a \cdot (p_1+p_2)}\epsilon^{I_1I_2I_3[I_4I_5I_6I_7|J}  (R_a)^{I_8]}\,_J M_{n-1}(g)\,.
\eqae
Thus after removing the auxiliary graviton via the soft-graviton theorem, and summing over all relevant BCFW channels, we obtain:
\eqa\label{GenSoft2}
\nonumber &&\hspace{-0.5cm}M_{n}\left[\psi^{I_1I_2I_3}(t^2 p_1),\,\bar\psi^{I_4I_5I_6I_7I_8} (t^2 p_2)\cdots \right]\bigg|_{t\rightarrow 0}\\
& &\hspace{-0.5cm}~~~~~=\sum_{a=3}^n5\mathcal{F}_a\epsilon^{I_1I_2I_3[I_4I_5I_6I_7| J}(R_a)^{I_8]}\,_J M_{n{-}2}\,,
\eqae
where $\mathcal{F}_a\equiv \frac{[1|p_a |2\rangle}{2p_a\cdot(p_1+p_2)}$. Note the remarkable similarity to the double-soft limit of scalars, here one simply replaces the factor $p_a\cdot(p_1{-}p_2)$ in the numerator of $\mathcal{B}_a$ with $[1|p_a |2\rangle$. 

\section{New double-soft theorems in three dimensions}  

Next we consider amplitudes in three-dimensional $\mathcal N=16$ supergravity~\cite{E8} with two soft fermions. Again the on-shell degrees of freedom, 128 bosons and 128 fermions, can be packaged into a superfield  
\eqa
\Phi=\xi+\sum_{n=1}^{8} \xi_{I_1\dots I_n}\eta^{I_1}\eta^{I_2}\cdots\eta^{I_n},~~I_i=1,\dots,8\,.
\eqae
The Grassmann variables $\eta^{I}$ transform as fundamentals of SU(8)$\subset$SO(16), thus only part of the SO(16) are linearly realized in these variables, with the remaining realized non-linearly. The 128 bosons parametrize the coset space E$_{8(8)}$/SO(16). \par
 
Due to the presence of U(1) in U(8), the double-soft limit is again polluted by soft-graviton divergences. Thus inspired by~\cite{US}, we consider the \textit{symmetrized} double-soft limit:
\eq
\mathcal{M}^{\{i,j\}}_{n}\equiv \mathcal{M}_n\left( \psi^i( t^2 p_1),\; \psi^j( t^2 p_2),\cdots,n \right)+(1\leftrightarrow 2)\bigg|_{t {\rightarrow}0}\,.
\eqe
Using the BCFW representation of three-dimensional supergravity amplitudes, we find that remarkably amplitudes with the symmetrized double-soft fermions also behave universally and are given by:
\eqa
\mathcal{M}^{\{i,j\}}_{n} =-\sum_{a=3}^{n} (\mathcal{S}_a)_{i,j}\mathcal{M}_{n-2}+\mathcal O(t^2)\,,
\eqae 
 where $(\mathcal{S}_a)_{i,j}$ are the corresponding soft factors acting on the $(n{-}2)$-point superamplitude, and are given by:\\[-1mm]
\eqa
\notag &&\hspace{-0.5cm}\begin{array}{lcl}
(\mathcal{S}_a)^{I_1\dots I_v}_{J_1\dots J_{v+2}}&=&\frac{(v+2)!\mathcal F_a}{2!(-1)^{\delta_{v,3}}}\delta^{[I_1\dots I_v]}_{[J_1\dots J_v}(R_a)_{J_{v+1}J_{v+2}]}\,,\\
(\mathcal{S}_a)^{I_1\dots I_{v+2}}_{J_1\dots J_v}&=&\frac{(v+2)!\mathcal F_a}{(-1)^{\delta_{v,3}}}\delta^{[I_1\dots I_{v+2}]}_{[J_1\dots J_v] IJ}(R_a)^{IJ}\,,\\
(\mathcal{S}_a)^{I_1\dots I_v}_{J_1\dots J_v}&=&\frac{v!\mathcal F_a}{(-1)^{\delta_{v,1}}}\delta_{I[J_1\dots J_{v-1}}^{[I_1\dots I_v]}[2v (R_a)^{I}_{J_v]}-\delta^I_{J_v]}R_a]\,,
\end{array}\\
&&~~ ~~~~\delta^{I_1\dots I_j}_{J_1\dots J_j}=\delta^{I_1}_{J_1}\delta^{I_2}_{J_2}\cdots\delta^{I_j}_{J_j}\,,~v=\left\{1,3\right\}\,.
\eqae
Here $(R_a)\equiv \eta_a^I\partial_{\eta_a^I}-4$, $(R_a)_{IJ}\equiv \partial_{\eta_a^I}\partial_{\eta_a^J}$, $(R_a)^{IJ}\equiv \eta_a^I\eta_a^J$ and $\mathcal{F}_a\equiv \frac{\langle 1|p_a |2\rangle}{2p_a\cdot (p_1+p_2)}$. Different $\mathcal{S}_a$s correspond to different choices of soft-fermion pairs. Note that although the soft-fermion limit behaves in a similar fashion as that of the bosons, its detailed algebra is different. This is reflected in the fact that they form distinct representations under the on-shell SU(8) symmetry.  
\section{New soft theorems at loop level}  
It is interesting to see if the new soft theorems in supergravity theories are subject to any loop corrections. We begin with $\mathcal{N}=16$ supergravity in three dimensions, whose one-loop amplitudes can be expressed in terms of scalar triangle integrals with coefficients determined by generalized unitarity cuts. We can then apply tree-level soft theorems since it is the tree-level amplitude that enters the cuts. Follow the same proof of their scalar partner~\cite{US}, it is straightforward to show that single- and double-soft-fermion theorems do not receive any one-loop corrections. For the theories in four dimensions, one can also express the amplitudes in terms of box integrals, which immediately shows that one-loop amplitudes with one soft fermion also vanish. However, for the double-soft limit, unlike the three-dimensional case where amplitudes only with even-number external legs exist, new complication arises due to the discontinuity of the integral functions~\cite{BTWsoft}. Here we provide some evidence that the new soft theorems are not corrected by loops by considering the leading IR-divergent part of a $L$-loop amplitude, which is given by:
\eq
M^{\rm L-loop}_n\big{|}_{\rm lead. IR} ={1 \over L!} 
\left(\sum^n_{i,j} {  s_{ij} \log(s_{ij}) \over \epsilon} \right)^L M^{\rm tree}_{n}\, .
\eqe
Thanks to the kinematics factor $s_{ij}$, applying the tree-level soft theorems to $M^{\rm tree}_{n}$ we find, at least, the leading IR-divergent part of loop amplitudes satisfies the same soft theorems as tree-level amplitudes. 

\section{Conclusions}  

In this letter, we propose new soft theorems by studying soft fermions for the amplitudes in a wide range of theories, including Akulov-Volkov theory and supergravity theories in four and three dimensions. We find that all the amplitudes vanish in the single-soft limit, and behave universally in the double-soft limit, analogous to soft Goldstone bosons. The results for Akulov-Volkov theory precisely reflect that it is the effective theory for spontaneously supersymmetry breaking. To our surprises, the double-soft-fermion limits in extended supergravity theories are also universal and mimic the behavior of the scalars. We also provide evidence that the soft theorems do not receive loop corrections. Finally we like to emphasize that the results do not follow from the combination of double-soft scalar limits and supersymmetric Ward identities. It would be of great interest to clarify the implications of all those new soft theorems, in particular whether there are new hidden symmetries behind them, and their possible application for constraining potential ultraviolet counter terms for supergravity theories.

\section{Acknowledgements}
We would like to thank Zohar Komargodski for suggestions in looking into the Akulov-Volkov theory. We would also like to thank Massimo Bianchi, Paul Howe and Henrik Johansson for useful communications. Y-t. Huang is supported by MOST under the grant
No. 103-2112-M-002-025-MY3. W-M. Chen is supported by MOST under the grant No. 103-2112-M-002-025-MY3 and No. 104-2917-I-002-014.


\vskip .3 cm 


\begin{thebibliography}{99}
 \bibitem{Weinberg} 
  S.~Weinberg,
  Phys.\ Rev.\  {\bf 140}, B516 (1965).
\bibitem{Low} 
  F.~E.~Low,
  Phys.\ Rev.\  {\bf 110}, 974 (1958).

 \bibitem{SoftGravy1}
  D.~J.~Gross and R.~Jackiw,
  Phys.\ Rev.\  {\bf 166}, 1287 (1968);\\
  R.~Jackiw,
  Phys.\ Rev.\  {\bf 168}, 1623 (1968).



\bibitem{BernGauge} 
  Z.~Bern, S.~Davies, P.~Di Vecchia and J.~Nohle,
  Phys.\ Rev.\ D {\bf 90}, no. 8, 084035 (2014)
  [arXiv:1406.6987 [hep-th]].

 
\bibitem{CS} 
  A.~Strominger,
  JHEP {\bf 1407}, 152 (2014)
  [arXiv:1312.2229 [hep-th]].
    T.~He, V.~Lysov, P.~Mitra and A.~Strominger,
  arXiv:1401.7026 [hep-th].
  F.~Cachazo and A.~Strominger,
  arXiv:1404.4091 [hep-th].

\bibitem{SymArg} 
  A.~J.~Larkoski,
  Phys.\ Rev.\ D {\bf 90}, 087701 (2014)
  [arXiv:1405.2346 [hep-th]];\\
  J.~Broedel, M.~de Leeuw, J.~Plefka and M.~Rosso,
  Phys.\ Rev.\ D {\bf 90}, 065024 (2014)
  [arXiv:1406.6574 [hep-th]].
  J.~Broedel, M.~de Leeuw, J.~Plefka and M.~Rosso,
  arXiv:1411.2230 [hep-th].
  
\bibitem{AdlerZero} 
  S.~L.~Adler,
  Phys.\ Rev.\  {\bf 137}, B1022 (1965).
\bibitem{Weinberg:1966kf} 
  S.~Weinberg,
  Phys.\ Rev.\ Lett.\  {\bf 17}, 616 (1966).
  
  
\bibitem{Simplest} 
  N.~Arkani-Hamed, F.~Cachazo and J.~Kaplan,
  JHEP {\bf 1009}, 016 (2010)
  [arXiv:0808.1446 [hep-th]].
\bibitem{ElvangR4} 
  H.~Elvang and M.~Kiermaier,
  JHEP {\bf 1010}, 108 (2010)
  [arXiv:1007.4813 [hep-th]].


\bibitem{ElvangFull} 
  N.~Beisert, H.~Elvang, D.~Z.~Freedman, M.~Kiermaier, A.~Morales and S.~Stieberger,
  Phys.\ Lett.\ B {\bf 694}, 265 (2010)
  [arXiv:1009.1643 [hep-th]].
  
  
 \bibitem{VAModel} 
  D.~V.~Volkov and V.~P.~Akulov,
  JETP Lett.\  {\bf 16}, 438 (1972)
  [Pisma Zh.\ Eksp.\ Teor.\ Fiz.\  {\bf 16}, 621 (1972)].
  
  
 \bibitem{E8} 
  N.~Marcus and J.~H.~Schwarz,
  Nucl.\ Phys.\ B {\bf 228}, 145 (1983).
  
 

\bibitem{US} 
  W.~M.~Chen, Y.~t.~Huang and C.~Wen,
  arXiv:1412.1811 [hep-th].




\bibitem{SUSYWard} 
  M.~T.~Grisaru, H.~N.~Pendleton and P.~van Nieuwenhuizen,
  Phys.\ Rev.\ D {\bf 15}, 996 (1977).



\bibitem{BCFW}
R.~Britto, F.~Cachazo, B.~Feng and E.~Witten,
  Phys.\ Rev.\ Lett.\  {\bf 94}, 181602 (2005)
  [hep-th/0501052];\\
R.~Britto, F.~Cachazo and B.~Feng,
  Nucl.\ Phys.\ B {\bf 715}, 499 (2005)
  [hep-th/0412308].
\bibitem{LessSUSY} 
  H.~Elvang, Y.~t.~Huang and C.~Peng,
  JHEP {\bf 1109}, 031 (2011)
  [arXiv:1102.4843 [hep-th]].

  
  


   
\bibitem{BTWsoft}
  A.~Brandhuber, G.~Travaglini and C.~Wen,
  work in progress.
  
\end{thebibliography}
\end{document}